\documentclass[12pt,a4paper]{article}
\usepackage{amsmath,amssymb,amsfonts,amsthm,amsopn,graphicx}
\usepackage{tikz} 
\usetikzlibrary{arrows}

\usepackage{xcolor}

\allowdisplaybreaks[3]
\newtheorem{theorem}{Theorem}[section]

\newtheorem{conjecture}[theorem]{Conjecture}

\newcommand{\ds}{\displaystyle}
\def\EXP{\textrm{{\large e}}}
\newcommand{\ii}{\mathsf{i}}
\newcommand{\mPsi}{X}
\newcommand{\mPhi}{Y}
\renewcommand{\author}[1]{\large\rm #1\\ \bigskip}
\newcommand{\address}[1]{{\normalsize\it #1\\}\bigskip}
\renewcommand{\title}[1]{\bigskip\bigskip\Large\bf #1\bigskip\bigskip\\}

\newcommand{\bch}{\boldsymbol{\chi}}
\newcommand{\Qop}{\mathcal{Q}}

\usepackage[body={15.5cm,22.5cm}]{geometry} 
\begin{document}
\vglue 2cm

\begin{center}

\title{Elliptic Sklyanin  Algebra, Baxter Equation and Discrete Liouville Equation.}
\author{Sergey M.~Sergeev.}

\vspace{.5cm}

\address{Department of Theoretical Physics,
         Research School of Physics and Engineering,\\
    Australian National University, Canberra, ACT 0200, Australia\\
    and\\
   Faculty of Science and Technology, \\
   University of Canberra, Bruce ACT 2617, Australia }


\end{center}
\begin{abstract}
In this paper we discuss some properties of Baxter's TQ equation for the eight-vertex elliptic Sklyanin algebra it its compact representation based on the elliptic Gamma-functions. As the main result, we establish the structure of the spectrum of the corresponding quantum integrable model.
\end{abstract}



\section{Introduction}

In this paper we consider the Baxter's TQ equation \cite{Baxterbook} for the elliptic Sklyanin algebra \cite{Skl82}. Namely, let $T(u)$ stands for a monodromy matrix for the Sklyanin algebra, i.e. it satisfies
\begin{equation}
R_{12}(x_1-x_2) T_1(x_1) T_2(x_2) \;=\; T_2(x_2) T_1(x_1) R_{12}(x_1-x_2)\;,
\end{equation}
where $R_{12}(x)$ is the eight-vertex $R$-matrix with elliptic modulus $\tau$ and crossing parameter $\eta$, and $T(x)$ is defined in specific representation related to the Baxter's vectors.  

The representation structure of $T(x)$ allows one to introduce the Ising-type weights and to define $\Qop$-operators. In the most general case the Ising-type weights are expressed in the terms of elliptic $\Gamma$-function \cite{Spiridonov-beta}, symmetric with respect to exchange $\tau\leftrightarrow \eta$. The commutativity of $\Qop$-operators is based on Spiridonov integral identity \cite{Spiridonov-beta} rewritten as the Star-Triangle relation \cite{BS:Master}, and we will refer to such Ising-type weights as to the Master solution of the Yang-Baxter equation. However, the definition of $\Qop$-operator as well as the Baxter TQ equation are based on simple algebraic properties of Baxter's vectors. 

The TQ equation in our case reads
\begin{equation}\label{TQ00}
t(x|\tau) \Qop(x) \;=\; \theta_1(x+y|\tau)^N \Qop(x+\eta) + \theta_1(x-y|\tau)^N \Qop(x-\eta)\;,
\end{equation}
where $t(x)\;=\;\textrm{tr} \;T(x)$ is the ``auxiliary'' transfer-matrix, and $y$ is an extra parameter of the representation not related to $\eta$ and $\tau$.

The main subject of this paper is equation (\ref{TQ00}) and its consequence, the discrete Liouville equation,
\begin{equation}\label{Liou00}
\Qop(x+\eta+\tau)\Qop(x) \;-\; \EXP^{2\pi\ii y N} \,\Qop(x+\eta) \Qop(x+\tau)\;=\;R(x,y)\;,
\end{equation}
where $R(x,y)$ is some (\emph{\`a priori} known) elliptic $\Gamma$-function. We study these equations in several regimes where solutions $\Qop(x)$ have different structures. This will be the subject of Section \ref{section3}.

However, before this, in Section \ref{section2} we give an exposition of the method of Baxter's vectors, eight-vertex ``Solid-on-Solid'' (SOS) model, representation of Sklyanin algebra, derivation of Ising-type weights and derivation of TQ equations.

\section{Baxter vectors for the eight-vertex model and Ising-type weights}\label{section2}

We remind the Reader the method of Baxter vectors for the eight-vertex model \cite{Baxterbook} in this section. See also \cite{Zabrodin}. The purpose of it is to fix the main reference points.

\subsection{Notations}

Firstly, let us fix all notations for elliptic theta-functions and elliptic $\Gamma$-functions.

Here we will use the standard definitions of $\theta$-functions,
\begin{equation}
\theta_1(x|\tau) \;=\; -\ii \sum_{n\in\mathbb{Z}} (-)^n \EXP^{\ii\pi\tau (n+1/2)^2 + 2\pi\ii (n+1/2) x}\;,
\end{equation}
and
\begin{equation}
\theta_4(x|\tau)\;=\;\sum_{n\in\mathbb{Z}} (-)^n \EXP^{\ii\pi\tau n^2 + 2\pi\ii n x}\;.
\end{equation}
Note, the quasi-periods of $\theta(x|\tau)$ are $1$ and $\tau$, so that $\theta_2(x|\tau)=\theta_1(x+\frac{1}{2}|\tau)$, $\theta_3(x|\tau) = \theta_4(x+\frac{1}{2}|\tau)$.

Through this paper we will use the following correspondence:
\begin{equation}
u\;=\;\EXP^{2\pi\ii x}\;,\quad 
v\;=\;\EXP^{2\pi\ii y}\;,\quad
q\;=\;\EXP^{2\pi\ii\tau}\;,\quad
p\;=\;\EXP^{2\pi\ii\eta}\;.
\end{equation}
Alternative expression for the shortened theta-function is
\begin{equation}
h_q(u)\;=\;h(u;q)\;=\;(u;q)_\infty (q/u;q)_\infty\;.
\end{equation}
In particular,
\begin{equation}
\theta_1(x|\tau)\;=\; \ii q^{1/8} u^{-1/2} h(u;q) (q;q)_\infty\;,\quad 
\theta_4(x|\tau)\;=\; h(q^{1/2}u;q) (q;q)_\infty\;.
\end{equation}

Elliptic $\Gamma$-function is defined by
\begin{equation}
\Gamma(u)\;=\;\frac{(pq/u;p,q)_\infty}{(u;p,q)_\infty}\;.
\end{equation}
Its main difference property is 
\begin{equation}
\frac{\Gamma(pu)}{\Gamma(u)}\;=\; h(u;q)\;.
\end{equation}
Often we will use the series decomposition of $\log \Gamma$,
\begin{equation}
\log \Gamma(u) \;=\; \sum_{n\neq 0} \frac{u^k}{k(1-q^k)(1-p^k)} \;,\quad pq<u<1\;.
\end{equation}
One more form of $\Gamma$-function is
\begin{equation}\label{PG1}
\Phi(x)\;=\;\frac{(\sqrt{pq}u;p,q)_\infty}{(\sqrt{pq}/u;p,q)_\infty}\;=\;\frac{1}{\Gamma(\sqrt{pq}u)}\;,
\end{equation}
satisfying
\begin{equation}\label{PG2}
\frac{\Phi(x-\frac{\eta}{2})}{\Phi(x+\frac{\eta}{2})}\;=\;h(q^{1/2}u;q)\;\sim\;\theta_4(x|\tau)\;.
\end{equation}

\subsection{Eight-vertex weights}

Let the elements of the eight-vertex $R$-matrix are given by
\begin{equation}\label{eight-vertex}
\begin{array}{c}
\ds R_{0,0}^{0,0}(x)\;=\;R_{1,1}^{1,1}(x)\;=\;\varrho \; \theta_1(x+\eta|2\tau) \theta_4(x|2\tau) \theta_4(\eta|2\tau)\;,\\
\\
\ds R_{0,1}^{0,1}(x)\;=\;R_{1,0}^{1,0}(x)\;=\;\varrho \; \theta_4(x+\eta|2\tau) \theta_1(x|2\tau) \theta_4(\eta|2\tau)\;,\\
\\
\ds R_{0,1}^{1,0}(x)\;=\;R_{1,0}^{0,1}(x)\;=\;\varrho \; \theta_4(x+\eta|2\tau) \theta_4(x|2\tau) \theta_1(\eta|2\tau)\;,\\
\\
\ds R_{0,0}^{1,1}(x)\;=\;R_{1,1}^{0,0}(x)\;=\;\varrho \; \theta_1(x+\eta|2\tau) \theta_1(x|2\tau) \theta_1(\eta|2\tau)\;.
\end{array}
\end{equation}

\subsection{Baxter vectros}

Baxter vectors are the subject of the Vertex-SOS duality:
\begin{equation}\label{Rpsi}
\begin{tikzpicture}
\draw [-open triangle 45,dashed] (2,3) -- (2,-1); 
\node [right] at (2.3,-1) {$a$};
\node [right] at (2.3,1) {$b$};
\node [right] at (2.3,3) {$c$};
\draw [-latex, thick] (2,0) -- (-1,1.5); \node[left] at (-1,1.5) {$x$};
\draw[dotted, thick] (2,0) -- (3,-0.5);
\draw [-latex, thick] (2,2) -- (-1,0.5); \node[left] at (-1,0.5) {$x'$};
\draw[dotted, thick] (2,2) -- (3,2.5);
\end{tikzpicture}
\begin{tikzpicture}
\node [right] at (-3,1) {$\ds =\quad \sum_d$};
\draw [-open triangle 45,dashed] (0,3) -- (0,-1); 
\node [right] at (2,-1) {$a$};
\node [right] at (3,1) {$b$};
\node [right] at (2,3) {$c$};
\node [right] at (0.3,1) {$d$};
\draw [-latex, thick] (0,0) -- (-1,-0.5); \node[left] at (-1,2.5) {$x$};
\draw[dotted, thick] (0,0) -- (3,1.5);
\draw [-latex, thick] (0,2) -- (-1,2.5); \node[left] at (-1,-0.5) {$x'$};
\draw[dotted, thick] (0,2) -- (3,0.5);
\end{tikzpicture}
\end{equation}
Here the ``solid'' lines correspond to two dimensional vector space labeled by $0,1$, the ``solid cross'' in the left hand side stands for the completely symmetric eight-vertex matrix $R(x-x')_{i_1,i_2}^{j_1,j_2}$, eq. (\ref{eight-vertex}). The other elements of (\ref{Rpsi}) are defined as follows. Bexter vectors,
%
%
%
\begin{equation}\label{Psi1}
\begin{tikzpicture}
\draw [-open triangle 45,dashed] (1.5,3) -- (1.5,0); \node [below] at (1.5,0) {$y$}; \node [above] at (1.5,3) {$\phantom{y}$};
\draw [-latex, thick] (1.5,1.5) -- (0,1.5); \node[left] at (0,1.5) {$x$};
\draw[dotted, thick] (1.5,1.5) -- (3,1.5); \node[right] at (2,2.5) {$a\mp\frac{\eta}{2}$}; \node[right] at (2,0.5) {$a$};
\node[right] at (3.3,1.5) {$\ds =\;\;\mPsi_a^{(\mp)}(x-y)\;,$};
\end{tikzpicture}
\quad
\begin{tikzpicture}
\draw [-open triangle 45,dashed] (1.5,0) -- (1.5,3); \node [above] at (1.5,3) {$y$}; \node [below] at (1.5,0) {$\phantom{y}$};
\draw [-latex, dotted, thick] (1.5,1.5) -- (0,1.5); \node[left] at (0,1.5) {$x$};
\draw[thick] (1.5,1.5) -- (3,1.5); \node[left] at (1,2.5) {$a\mp\frac{\eta}{2}$}; \node[left] at (1,0.5) {$a$};
\node[right] at (3.3,1.5) {$\ds =\;\;\overline{\mPsi}^{(\mp)}_a(x-y)\;,$};
\end{tikzpicture}
\end{equation}
\begin{equation}\label{Psi2}
\begin{tikzpicture}
\draw [-open triangle 45,dashed] (1.5,0) -- (1.5,3); \node [above] at (1.5,3) {$y$}; \node [below] at (1.5,0) {$\phantom{y}$};
\draw [-latex, thick] (1.5,1.5) -- (0,1.5); \node[left] at (0,1.5) {$x$};
\draw[dotted, thick] (1.5,1.5) -- (3,1.5); \node[right] at (2,2.5) {$a$}; \node[right] at (2,0.5) {$a\mp\frac{\eta}{2}$};
\node[right] at (3.3,1.5) {$\ds =\;\;\mPhi^{(\mp)}_a(x-y)\;,$};
\end{tikzpicture}
\quad
\begin{tikzpicture}
\draw [-open triangle 45,dashed] (1.5,3) -- (1.5,0); \node [above] at (1.5,3) {$\phantom{y}$}; \node [below] at (1.5,0) {$y$};
\draw [-latex, dotted, thick] (1.5,1.5) -- (0,1.5); \node[left] at (0,1.5) {$x$};
\draw[thick] (1.5,1.5) -- (3,1.5); \node[left] at (1,2.5) {$a$}; \node[left] at (1,0.5) {$a\mp\frac{\eta}{2}$};
\node[right] at (3.3,1.5) {$\ds =\;\;\overline{\mPhi}_a^{(\mp)}(x-y)\;.$};
\end{tikzpicture}
\end{equation}
where
\begin{equation}\label{Psi}
\mPsi_a^{(\mp)}(x)\;=\;\left(\begin{array}{c}
\ds \theta_1(x\pm 2a|2\tau) \\ \\ \ds \theta_4(x\pm 2a|2\tau)
\end{array}\right)\;,\quad
\overline{\mPsi}_a^{\;(\mp)}(x)\;=\;
\frac{\biggl(\pm\theta_4(x\mp 2a|2\tau),\mp\theta_1(x\mp 2a|2\tau)\biggr)}{\theta_1(2a|\tau)}\;,
\end{equation}
and
\begin{equation}\label{Phi}
\mPhi_a^{\;(\mp)}(x)\;=\;\left(\begin{array}{cc} 0 & 1 \\ -1 & 0 \end{array}\right) \mPsi^{\;(\mp)}_a(-x)\;,\quad
\overline{\mPhi}^{\;(\mp)}_a(x)\;=\;\overline{\mPsi}^{\;(\mp)}_a(-x) \left(\begin{array}{cc} 0 & -1 \\ 1 & 0 \end{array}\right)\;.
\end{equation}
Note the inversion relation,
\begin{equation}\label{inv}
\overline{\mPsi}_a^{(\epsilon)}(x) \mPsi_a^{(\epsilon')}(x)\;=\;\delta_{\epsilon,\epsilon'}\theta_2(x|\tau)\;,
\end{equation}
and the same for $\overline{\mPhi},\mPhi$.

\subsection{``Solid-on-Solid'' weights}

Eight-vertex SOS weights are then given by
\begin{equation}\label{SOS-a-I}
\begin{tikzpicture}
\draw [-latex, dotted, thick] (1,1) -- (-1,-1); 
\draw [-latex, dotted, thick] (1,-1) -- (-1,1); 
\node [above] at (-1,1) {$x$};
\node [below] at (-1,-1) {$x'$};
\node [left] at (-1,0) {$a$};
\node [below] at (0,-1) {$a-\frac{\eta}{2}$};
\node [above] at (0,1) {$a+\frac{\eta}{2}$};
\node [right] at (1,0) {$a$};
\node [right] at (1.7,0) {$=$};
\draw [-latex, dotted, thick] (5,1) -- (3,-1); 
\draw [-latex, dotted, thick] (5,-1) -- (3,1); 
\node [above] at (3,1) {$x$};
\node [below] at (3,-1) {$x'$};
\node [left] at (3,0) {$a$};
\node [below] at (4,-1) {$a+\frac{\eta}{2}$};
\node [above] at (4,1) {$a-\frac{\eta}{2}$};
\node [right] at (5,0) {$a$};
\node [right] at (6,0) 
{$\ds =\;\;\;\;\varrho' \;\theta_1(x-x'+\eta|\tau)$};
\end{tikzpicture}
\end{equation}
\begin{equation}\label{SOS-b-I-1}
\begin{tikzpicture}
\draw [-latex, dotted, thick] (1,1) -- (-1,-1); 
\draw [-latex, dotted, thick] (1,-1) -- (-1,1); 
\node [above] at (-1,1) {$x$};
\node [below] at (-1,-1) {$x'$};
\node [left] at (-0.5,0) {$a-\frac{\eta}{2}$};
\node [below] at (0,-1) {$a$};
\node [above] at (0,1) {$a$};
\node [right] at (0.5,0) {$a+\frac{\eta}{2}$};
\node [right] at (2,0) 
{$\ds =\;\;\;\;\ds \varrho' \;\frac{\theta_1(x-x'|\tau)\theta_1(2a+\eta|\tau)}{\theta_1(2a|\tau)}$};
\end{tikzpicture}
\end{equation}
\begin{equation}\label{SOS-b-I-2}
\begin{tikzpicture}
\draw [-latex, dotted, thick] (1,1) -- (-1,-1); 
\draw [-latex, dotted, thick] (1,-1) -- (-1,1); 
\node [above] at (-1,1) {$x$};
\node [below] at (-1,-1) {$x'$};
\node [left] at (-0.5,0) {$a+\frac{\eta}{2}$};
\node [below] at (0,-1) {$a$};
\node [above] at (0,1) {$a$};
\node [right] at (0.5,0) {$a-\frac{\eta}{2}$};
\node [right] at (2,0) 
{$\ds =\;\;\;\;\ds \varrho'\;\frac{\theta_1(x-x'|\tau)\theta_1(2a-\eta|\tau)}{\theta_1(2a|\tau)}$};
\end{tikzpicture}
\end{equation}
\begin{equation}\label{SOS-c-I-1}
\begin{tikzpicture}
\draw [-latex, dotted, thick] (1,1) -- (-1,-1); 
\draw [-latex, dotted, thick] (1,-1) -- (-1,1); 
\node [above] at (-1,1) {$x$};
\node [below] at (-1,-1) {$x'$};
\node [left] at (-0.5,0) {$a+\frac{\eta}{2}$};
\node [below] at (0,-1) {$a$};
\node [above] at (0,1) {$a$};
\node [right] at (0.5,0) {$a+\frac{\eta}{2}$};
\node [right] at (2,0) 
{$\ds =\;\;\;\;\varrho'\;\frac{\theta_1(\eta|\tau)\theta_1(x-x'-2a|\tau)}{\theta_1(2a|\tau)}$};
\end{tikzpicture}
\end{equation}
\begin{equation}\label{SOS-c-I-2}
\begin{tikzpicture}
\draw [-latex, dotted, thick] (1,1) -- (-1,-1); 
\draw [-latex, dotted, thick] (1,-1) -- (-1,1); 
\node [above] at (-1,1) {$x$};
\node [below] at (-1,-1) {$x'$};
\node [left] at (-0.5,0) {$a-\frac{\eta}{2}$};
\node [below] at (0,-1) {$a$};
\node [above] at (0,1) {$a$};
\node [right] at (0.5,0) {$a-\frac{\eta}{2}$};
\node [right] at (2,0) 
{$\ds =\;\;\;\;\varrho'\; \frac{\theta_1(\eta|\tau)\theta_1(x-x'+2a|\tau)}{\theta_1(2a|\tau)}$};
\end{tikzpicture}
\end{equation}

\subsection{$L$-operators}

One can construct several $L$-operators:
\begin{equation}\label{LL}
\begin{tikzpicture}
\draw [-open triangle 45, dashed, thick] (1.5,1.5) -- (1.5,-1.5); \node [below] at (1.5,-1.5) {$y_1$};
\draw [-open triangle 45, dashed, thick] (2.5,-1.5) -- (2.5,1.5); \node [above] at (2.5,1.5) {$y_2$};
\draw [-latex, thick] (1.5,0) -- (0,0); \node [left] at (0,0) {$x$};
\draw [thick, dotted] (2.5,0) -- (1.5,0); 
\draw [thick] (4,0) -- (2.5,0);
\node [below] at (2,-1) {$a$};
\node [above] at (2,1) {$b$};
\node [below] at (2,-2.5) {$\langle b| L(x) |a\rangle$};
\end{tikzpicture}
\qquad\qquad
\begin{tikzpicture}
\draw [-open triangle 45, dashed, thick] (1.5,-1.5) -- (1.5,1.5); \node [above] at (1.5,1.5) {$y_2$};
\draw [-open triangle 45, dashed, thick] (2.5,1.5) -- (2.5,-1.5); \node [below] at (2.5,-1.5) {$y_1$};
\draw [-latex, thick] (1.5,0) -- (0,0); \node [left] at (0,0) {$x$};
\draw [thick, dotted] (2.5,0) -- (1.5,0); 
\draw [thick] (4,0) -- (2.5,0);
\node [below] at (2,-1) {$a$};
\node [above] at (2,1) {$b$};
\node [below] at (2,-2.5) {$\langle b| L'(x) |a\rangle$};
\end{tikzpicture}
\end{equation}
or, in matrix elements,
\begin{equation}
\begin{array}{c}
\ds 
\langle a+\epsilon\frac{\eta}{2}| L(x) | a\rangle \;=\; \mPsi_a^{(\epsilon)}(x-y_1) \overline{\mPsi}_a^{(\epsilon)}(x-y_2)\;,\\
\\
\ds \langle b | L'(x) | b+\epsilon\frac{\eta}{2}\rangle \;=\; \mPhi_b^{(\epsilon)}(x-y_2) \overline{\mPhi}_b^{(\epsilon)}(x-y_1)\;.
\end{array}
\end{equation}
It is convenient to get rid of the manifest difference property and to put
\begin{equation}
y_1=y\;,\quad y_2=-y\;.
\end{equation}
Then one can decompose $T$-matrices as follows:
\begin{equation}\label{Mop}
\begin{tikzpicture}
\draw [-open triangle 45, dashed, thick] (1.5,-1.5) -- (1.5,1.5); \node [above] at (1.5,1.5) {$-y$};
\draw [-open triangle 45, dashed, thick] (2.5,1.5) -- (2.5,-1.5); \node [below] at (2.5,-1.5) {$y$};
\draw [-latex, dotted, thick] (1.5,0) -- (0,0); \node [left] at (0,0) {$x$};
\draw [thick] (2.5,0) -- (1.5,0); 
\draw [dotted, thick] (4,0) -- (2.5,0);
\node [left] at (1,-1) {$a$};
\node [left] at (1,1) {$a+\epsilon\frac{\eta}{2}$};
\node [right] at (3,-1) {$a'$};
\node [right] at (3,1) {$a'+\epsilon'\frac{\eta}{2}$};
\node [right] at (5,0) 
{$
\ds \;=\; \frac{\theta_2(x+\epsilon a - \epsilon'a'|\tau) \theta_1(y+\epsilon a + \epsilon'a'|\tau)}{\theta_1(2\epsilon a|\tau)}\;,
$};
\end{tikzpicture}
\end{equation}
so that (\ref{Mop}) is rather $L$-operators for SOS weights. The matrix elements of the transfer matrix are
\begin{equation}\label{Top}
\langle a+\epsilon\frac{\eta}{2}| T(x) |a\rangle \;=\; \prod_{k=1}^N 
\frac{\theta_2(x+\epsilon_ka_k-\epsilon_{k+1}a_{k+1}|\tau)\theta_1(y+\epsilon_ka_k+\epsilon_{k+1}a_{k+1}|\tau)}{\theta_1(2\epsilon_ka_k|\tau)}\;.
\end{equation}
The second $L$-operator, $L'$, provides
\begin{equation}\label{Mopp}
\begin{tikzpicture}
\draw [-open triangle 45, dashed, thick] (1.5,1.5) -- (1.5,-1.5); \node [below] at (1.5,-1.5) {$y$};
\draw [-open triangle 45, dashed, thick] (2.5,-1.5) -- (2.5,1.5); \node [above] at (2.5,1.5) {$-y$};
\draw [-latex, dotted, thick] (1.5,0) -- (0,0); \node [left] at (0,0) {$x$};
\draw [thick] (2.5,0) -- (1.5,0); 
\draw [dotted, thick] (4,0) -- (2.5,0);
\node [left] at (1,-1) {$a+\epsilon\frac{\eta}{2}$};
\node [left] at (1,1) {$a$};
\node [right] at (3,-1) {$a'+\epsilon'\frac{\eta}{2}$};
\node [right] at (3,1) {$a'$};
\node [right] at (5,0) 
{$
\ds \;=\; \frac{\theta_2(x-\epsilon a + \epsilon'a'|\tau) \theta_1(y+\epsilon a + \epsilon'a'|\tau)}{\theta_1(2\epsilon a|\tau)}\;,
$};
\end{tikzpicture}
\end{equation}
giving the following matrix elements of the second transfer matrix:
\begin{equation}\label{Topp}
\langle a| T'(x) |a+\epsilon\frac{\eta}{2}\rangle \;=\; \prod_{k=1}^N 
\frac{\theta_2(x-\epsilon_ka_k+\epsilon_{k+1}a_{k+1}|\tau)\theta_1(y+\epsilon_ka_k+\epsilon_{k+1}a_{k+1}|\tau)}{\theta_1(2\epsilon_ka_k|\tau)}\;.
\end{equation}
Relation between two types of the transfer-matrices is straightforward:
\begin{equation}
T'(x) \;=\; T(-x)^t\;.
\end{equation}

\subsection{Ising-type weights}

Archetypical intertwining relation, defining the Ising-type weights, graphically is
\begin{equation}\label{W-1}
\begin{tikzpicture}
\draw [-open triangle 45, dashed] (-1.5,-1) -- (0.5,3) ; \node [above] at (0.5,3) {$x_1$};
\draw [-open triangle 45, dashed] (1.5,-1) -- (-0.5,3); \node [above] at (-0.5,3) {$x_2$};
\draw [-latex, thick] (1,0) -- (0,0); \draw [thick] (0,0) -- (-1,0); \node [left] at (-2,0) {$x$};
\draw[-latex, dotted, thick] (-1,0) -- (-2,0);  \draw[dotted, thick] (1,0) -- (2,0); 
\draw [fill] (1,2) circle [radius=0.07]; \draw [fill] (-1,2) circle [radius=0.07];
\draw [red, ultra thick] (1,2) -- (-1,2);
\node [left] at (-1,2) {$a'$}; \node[right] at (1,2) {$b'$};
\end{tikzpicture}
\quad\quad
\begin{tikzpicture}
\node [right] at (-4,1) {$=$};
\draw [-open triangle 45,dashed] (-0.5,-1) -- (1.5,3); \node [above] at (1.5,3) {$x_1$};
\draw [-open triangle 45,dashed] (0.5,-1) -- (-1.5,3); \node [above] at (-1.5,3) {$x_2$};
\draw [-latex, thick] (1,2) -- (0,2); \draw [thick] (0,2) -- (-1,2);
\draw[-latex, dotted, thick] (-1,2) -- (-2,2);  \draw[dotted, thick] (1,2) -- (2,2); \node [left] at (-2,2) {$x$};
\draw [fill] (1,0) circle [radius=0.07]; \draw [fill] (-1,0) circle [radius=0.07];
\draw [red, ultra thick] (1,0) -- (-1,0);
\node [left] at (-1,0) {$a$}; \node[right] at (1,0) {$b$};
\end{tikzpicture}
\end{equation} 

\noindent
The same relation in the precise notations is
\begin{equation}\label{arch}
\begin{array}{l}
\ds 
V_{x_2-x_1}(a+\epsilon\frac{\eta}{2},b') \overline{\mPsi}^{(\epsilon)}_a(x-x_1) \mPhi^{(\epsilon')}_{b'}(x-x_2) \;=\;\\
\\
\ds \phantom{xxxxxxxxxxx}\;=\;
\overline{\mPsi}^{(\epsilon)}_a(x-x_2) \mPhi^{(\epsilon')}_{b'}(x-x_1) V_{x_2-x_1}(a,b'+\epsilon'\frac{\eta}{2})\;.
\end{array}
\end{equation}
For rational $\eta$ this equation defines the Boltzmann weights for the Kashiwara-Miwa model \cite{Kashiwara:1986,Boos:1997}. However, the compact integral representation of Sklyanin algebra implies the elliptic $\Gamma$-function solution of (\ref{arch}):
\begin{equation}
V_x(a,b)\;=\;\frac{\Phi(a-b+\frac{x}{2}) \Phi(a+b+\frac{x}{2})}{\Phi(a-b-\frac{x}{2})\Phi(a+b-\frac{x}{2})}\;,
\end{equation}
where $\Phi$ is defined by (\ref{PG1}). Notations (\ref{Top}) and (\ref{Topp}) should be evidently equipped by the $\delta$-functions for the measure $\ds \int_0^1 da$.

The saw-type transfer matrix is defined graphically as follows:
\begin{equation}\label{P1}
\begin{tikzpicture}
\draw [-open triangle 45, dashed] (4.5,1) -- (-0.5,1); \node [left] at (-0.5,1) {$x$};
\draw [-open triangle 45, dashed] (1,2) -- (1,0); \node [below] at (1,0) {$y$};
\draw [-open triangle 45, dashed] (3,0) -- (3,2); \node [above] at (3,2) {$-y$};
\draw [fill] (0,2) circle[radius=0.07]; \node [above] at (0,2) {$a_k$};
\draw [red, ultra thick] (0,2) -- (2,0);
\draw [fill] (2,0) circle[radius=0.07]; \node [below] at (2,0) {$b_k$};
\draw [red, ultra thick] (2,0) -- (4,2);
\draw [fill] (4,2) circle[radius=0.07]; \node [above] at (4,2) {$a_{k+1}$};
\draw [red, ultra thick] (0,2) -- (-0.5,1.5);
\draw [red, ultra thick] (4,2) -- (4.5,1.5);
\node [right] at (5.5,1) 
{$\ds =\;\;\;\langle a | \Qop(x) | b \rangle$\;.};
\end{tikzpicture}
\end{equation}
In the explicit notations the kernel is given by
\begin{equation}
\langle a |\Qop(x) |b\rangle \;=\; \prod_{k=1}^N V_{y-x}(a_k,b_k) V_{y+x}(b_k,a_{k+1})\;.
\end{equation}
Intertwining relations, equations (\ref{W-1}) or (\ref{arch}), provide
\begin{equation}
\Qop(x) T(x') \;=\; T'(x') \Qop(x)\;.
\end{equation}
Normalised $\Qop$-s give the commutative family,
\begin{equation}\label{norm1}
\boldsymbol{Q}(x)\;=\;\Qop(x_0)^{-1}\Qop(x)\;,
\end{equation}
\begin{equation}
[T(x),T(x')]\;=\;[T(x),\boldsymbol{Q}(x')]\;=\;[\boldsymbol{Q}(x),\boldsymbol{Q}(x')]\;=\;0\;.
\end{equation}

\noindent
Degeneration point for Baxter's vectors is $x=1/2$, see (\ref{inv}). It is convenient to change the normalisation of the transfer matrix, 
\begin{equation}
T(x+\frac{1}{2}) \;=\; (-)^N t(x|\tau)\;.
\end{equation}
The projection matrix decomposition of Baxter's vectors provides the Baxter equation 
\begin{equation}\label{TQ0}
\Qop(x) t(x|\tau) \;=\; \theta_1(x+y|\tau)^N \Qop(x+\eta) \;+\; \theta_1(x-y|\tau)^N \Qop(x-\eta)\;.
\end{equation}
The second Baxter equation corresponds to exchange $\tau\leftrightarrow\eta$ what is the symmetry of $\Qop(x)$.

\section{Baxter equation}\label{section3}

\subsection{General considerations for $v<1$.}

Now it is more convenient to proceed to variables $u,v,q,p$. Let
\begin{equation}
\Qop(x)\; = \; \bch(u)\;,\quad t(x|\tau)\;=\; \left(\frac{\ii q^{1/8}(uv)^{-1/2}}{(q;q)_\infty}\right)^N \boldsymbol{t}_q(u)\;.
\end{equation}
Baxter' TQ equations can be rewritten as
\begin{equation}\label{TQ1}
\left\{\begin{array}{c}
\ds 
\boldsymbol{t}_q(u) \bch(u) \;=\; h_q(uv)^N \bch(pu) \;+\; v^N h_q(u/v)^N \bch(u/p)\;,\\
\\
\ds
\boldsymbol{t}_p(u) \bch(u) \;=\; h_p(uv)^N \bch(qu) \;+\; v^N h_p(u/v)^N \bch(u/q)\;.
\end{array}\right.
\end{equation}
Note e.g. the symmetries of the transfer-matrix,
\begin{equation}\label{tsym}
\boldsymbol{t}_q(qu)\;=\;\boldsymbol{t}_q(1/u)\;=\;(-u)^{-N} \boldsymbol{t}_q(u)\;,
\end{equation}
so that the transfer-matrix allows for instance the following decomposition:
\begin{equation}
\boldsymbol{t}_q(u)\;=\;\sum_{k=0}^{N-1} t_{q,k} h_q(\omega^k u)^N\;,\quad \omega\;=\;\EXP^{\frac{2\pi\ii}{N}}\;,\quad
t_{q,N-k}\;=\;t_{q,k}\;,
\end{equation}
so there are $\lfloor \frac{N}{2} \rfloor +1$ independent ``integrals of motion''.
Taking into account (\ref{tsym}) and considering the Wronskian of $\bch(u)$ and $\bch(qu)$ for the first of equations (\ref{TQ1}), one obtains the manifestly symmetric result:
\begin{equation}\label{Liouville}
\bch(pqu)\bch(u)\;-\;v^N \bch(pu)\bch(qu)\;=\;R(u)\;,\quad
R(u)\;=\;R_0 \left(\frac{\Gamma(1/uv)}{\Gamma(v/u)}\right)^N\;.
\end{equation}
This is the discrete Liouville equation since the right hand side satisfies the discrete Laplace equation
\begin{equation}
R(pqu) R(u) - v^{2N} R(pu) R(qu) \;=\; 0\;.
\end{equation}

Now we turn to the structure of solution of (\ref{TQ1}) and/or (\ref{Liouville}). So far we can suggest only perturbative approach, so our consideration can be seen as the existence and classification statements. One can choose any of (\ref{TQ1}) or (\ref{Liouville}) as the staring point, all three equations seem to be equivalent. To our opinion, the Liouville equation provides the most elegant scenario.

\bigskip

As the first step, take into account the poles of $\bch(u)$ and introduce $H(u)$,
\begin{equation}
\bch(u)\;=\;\frac{\ds H(u)}{\ds (\frac{pq}{uv},pq\frac{u}{v};p,q)_\infty}\;.
\end{equation}
Function $H(u)$ is convergent everywhere in $\mathbb{C}\setminus \{0,\infty\}$. Baxter's equation for $H(u)$ reads
\begin{equation}
\boldsymbol{t}_p(u) H(u) \;=\; (uv;p)_\infty^N (pq\frac{u}{v};p)_\infty^N H(qu) 
+ (-u)^N (\frac{v}{u};p)_\infty^N (\frac{pq}{uv};p)_\infty^N H(u/q)\;,
\end{equation}
and the Liouville equation becomes
\begin{equation}\label{Liou2}
H(pqu) H(u) - v^N (1-\frac{1}{uv})^N (1-pq\frac{u}{v})^N H(pu) H(qu) \;=\; S(u)\;,
\end{equation}
where
\begin{equation}
S(u) \;=\; R_0 \; (\frac{v}{u}, p^2q^2 \frac{u}{v}, pq uv, \frac{pq}{uv}; p,q)_\infty^N\;.
\end{equation}

\bigskip

Next, let symbolically
\begin{equation}
p\; \to \;\varepsilon p\;,\quad q\; \to \;\varepsilon q\;.
\end{equation}
Then the ``algebraic'' ground state is given by the series expansion,
\begin{equation}
H(u)\;=\;1 + \sum_{n=1}^\infty \varepsilon^{Q_n} H_n (u^n+u^{-n})\;,
\end{equation}
where $Q_n$ is some quadratic sequence\footnote{$Q=\{2,5,8,12,16,\dots\}$} with $Q_1=2$, and
\begin{equation}\label{HSexp}
H_n\;=\;\sum_{j=0}^\infty \varepsilon^j H_{n}^{(j)}\;,\quad
R_0\;=\;\sum_{j=0}^\infty \varepsilon^j R_{0}^{(j)}\;.
\end{equation}
All coefficients $H_{n}^{(j)}$ and $R_{0}^{(j)}$ can be obtained step-by-step from (\ref{Liou2}). It is important to note that the recursion is linear. Presumably, the seria (\ref{HSexp}) are convergent for $v<1$.

\bigskip

For the excited states, consider the set of zeros of the following simple polynomial:
\begin{equation}\label{pol-1}
G_m(u)\;=\;u^m (1-\frac{v}{u})^N + \frac{v^N}{u^m} (1-\frac{1}{uv})^N\;=\;\frac{1}{u^{m+N}}\prod_{j=1}^{m+N/2} (u-\xi_j^{})(u-\xi_j^{-1})\end{equation} 
for even $N$, and
\begin{equation}\label{pol-2}
G_m(u)\;=\;\frac{1}{u^{m+N}}(u-1)\prod_{j=1}^{m+(N-1)/2} (u-\xi_j^{})(u-\xi_j^{-1})
\end{equation} 
for odd $N$. Then, one has to separate the set of $\xi_j$ into two subsets (this is the classification step, in fact this is the solution of the spectral problem) and define
\begin{equation}\label{pol-3}
P^{(0)}_m(u) \;=\; \frac{1}{u^m} \prod_{j=1}^m (u-\xi_j^{})(u-\xi_j^{-1})\;=\; u^{-m} + \cdots + u^m\;,
\end{equation}
and
\begin{equation}\label{pol-4}
H^{(0)}(u)\;=\;\frac{1}{u^N} \prod_{j=m+1}^{m+N/2} (u-\xi_j^{})(u-\xi_j^{-1})\quad \textrm{or}\quad \frac{1}{u^N} (u-1) \prod_{j=m+1}^{m+(N-1)/2} (u-\xi_j^{})(u-\xi_j^{-1})\;.
\end{equation}
Polynomial $H^{(0)}$ has the decomposition
\begin{equation}
H^{(0)}(u)\;=\; 1 + \frac{H_{1}^{(0)}}{u} + \cdots + \frac{H^{(0)}_{N}}{u^N}\;.
\end{equation}
Polynomials $P^{(0)}_m$ and $H^{(0)}$ is the ``seed'' for $\varepsilon$-expansion
\begin{equation}
H(u) = P_m(u) + \varepsilon H_1(u^{m+1}+u^{-m-1}) + \cdots + \varepsilon^N H_N (u^{m+N}+u^{-m-N}) + o(\varepsilon^{N+1})\;,
\end{equation}
where there is a gap between $\varepsilon^N$ and higher powers. Expansion for $R_0$ must be modified,
\begin{equation}\label{R0m}
R_0\;=\; \sum_{j=0}^\infty \varepsilon^{j-m} R_0^{(j)}\;,
\end{equation}
while $\varepsilon$-expansion of $H_k$ and coefficients of $P_m$ start from $\varepsilon^0$ as in (\ref{HSexp}). The Liouville equation (\ref{Liou2}) again provides the step-by-step linear determination of all expansions.

\bigskip

Polynomials (\ref{pol-1},\ref{pol-2}) and (\ref{pol-3}) have simple meaning. This is just a solution of ``tropical'' TQ equation common for both equations (\ref{TQ1}) in the limit $p,q\to 0$:
\begin{equation}
\boldsymbol{t}^{(-m)}(u) P_m^{(0)}(u) \;=\; (1-uv)^N u^{-m} + (v-u)^N u^m \;=\; (-u)^N G_m(u)\;.
\end{equation}
Clear advantage of the ``tropical'' equation is that the combinatorics of the Bethe Ansatz roots becomes trivial.

\bigskip

Note, here we discussed the ``algebraic'' ground state. The integer $m$ in (\ref{pol-1}--\ref{pol-4}) is an analogue of total spin for a spin chain.
\begin{conjecture}  When the normalisation (\ref{norm1}) is taken into account, the eigenvalues of the operator $\boldsymbol{Q}$ are equivalent to
\begin{equation}
\boldsymbol{Q} \; = \; \frac{\bch(u)}{\sqrt{R_0}}\;\sim\;\varepsilon^{m/2}\;,
\end{equation}
so that the ``algebraic'' ground state and its excitations are the physical ground state and physical excitations.
\end{conjecture}

\bigskip

Consider now the Liouville equation for the ground state in the regime 
\begin{equation}\label{regime1}
\sqrt{pq} < v < 1\;,\quad \frac{pq}{v} < u < \frac{v}{pq}\;.
\end{equation}
Let
\begin{equation}
\log \frac{\bch(u)}{R_0} \;=\;\sum_{k=1}^\infty \; f_k \; (u^k+u^{-k})\;.
\end{equation}
Let next
\begin{equation}
f_k^{}\;=\;f_k^{(g)} + \delta f_k^{}\;,
\end{equation}
where
\begin{equation}
f_k^{(g)}\;=\;N \frac{(v^{-k}-v^k) p^k q^k}{k (1-p^k) (1-q^k) (1+p^kq^k)}\;.
\end{equation}
Such $f_k^{(g)}$ is the solution of $\bch^{(g)}(pqu) \bch^{(g)}(u)=R(u)$. Then, dividing (\ref{Liouville}) by $\bch(pqu)\bch(u)$ and taking the logarithm, one obtains the following equation for $\delta f_k$,
\begin{equation}
-\delta F_1(u) \;=\; \log (1- \digamma(u)^N \EXP^{-\delta F_2(u)}) \;,
\end{equation}
where
\begin{equation}
\delta F_1(u) \;=\; \sum_{k=1}^\infty \delta f_k \frac{1+k^kq^k}{p^kq^k} (u^{-k}+(pqu)^k)\;,
\end{equation}
\begin{equation}
\delta F_2(u) \;=\; \sum_{k=1}^\infty \delta f_k \frac{(1-p^k)(1-q^k)}{p^kq^k} (u^{-k}+(pqu)^k)\;,
\end{equation}
and
\begin{equation}
\digamma(u)\;=\; v \frac{h(1/uv;p^2q^2) h(pqu/v;p^2q^2)}{h(v/u;p^2q^2) h(pquv;p^2q^2)}\;.
\end{equation}
In the regime (\ref{regime1}) $|\digamma(u)|<1$, so that $\delta f_k\;\to\;0$ exponentially, and therefore $\bch^{(g)}(u)$ gives the expression for the free energy:
\begin{equation}\label{fren}
\frac{1}{N}\log \boldsymbol{Q}(x)\;=\;\sum_{k\neq 0} \frac{p^kq^k}{k (1-p^k)(1-q^k)(1+p^kq^k)} \left( (uv)^{-k} + (v/u)^{-k}\right)\;,
\end{equation}
explicitly in accordance with the results from \cite{BS:Master}.

\subsection{Special set for the case $v>1$.}

Perturbation expansion from the previous subsection shows an interesting feature. Function $\bch$ (or $H$)  in general position has denominators of the type $(1-p^mq^nv^2)^{-1}$.

It gives a special set of values of parameter $y$,
\begin{equation}
y\;=\;-\lambda\;,\quad \lambda\;=\; \frac{m}{2}\eta + \frac{n}{2}\tau\;.
\end{equation}
There is a temptation to call these set as $\ds (\frac{m}{2},\frac{n}{2})$-spin set.
In this case $R(u)$ has no denominator. Moreover, solution to (\ref{TQ1}) is given by
\begin{equation}
\Qop(x)\;=\;A(x|\tau)A'(x|\eta) + B(x|\tau) B'(x|\eta)\;,
\end{equation}
where
\begin{equation}
A(x|\tau)\;=\;\prod_{j=1}^{mN} \theta_4(\frac{x-x_j^{}}{2}|\frac{\tau}{2})\;,\quad 
A'(x|\eta)\;=\;\prod_{j=1}^{nN} \theta_4(\frac{x-x_j'}{2}|\frac{\eta}{2})\;,
\end{equation}
and $B(x|\tau)\;=\;A(x+1|\tau)$, $B'(x|\eta)=A'(x+1|\eta)$. Due to the requirement $A(-x)=A(x)$, the sets $\{x_j\}$ and $\{x_j'\}$ are symmetric. These sets are subject of the Bethe-Ansatz-type equations,
\begin{equation}
\left(\frac{\ds \theta_{\sigma(n)}(\frac{m}{2}\eta-x_k|\tau)}{\ds \theta_{\sigma(n)}(\frac{m}{2}\eta+x_k|\tau)}\right)^N
\prod_{j=1}^{mN} \frac{\ds \theta_1(\frac{x_k-x_j+\eta}{2}|\frac{\tau}{2})}{\ds \theta_1(\frac{x_k-x_j-\eta}{2}|\frac{\tau}{2})}\;=\;-1\;,
\end{equation}
where 
\begin{equation}
\sigma(\textrm{even})\;=\;4\;,\quad 
\sigma(\textrm{odd})\;=\;1\;,
\end{equation}
and symmetric equations with $\eta\leftrightarrow \tau$, $n\leftrightarrow m$, $x_j\leftrightarrow x_j'$. Note that in the Baxter's classical case $n=0$ and $m=1$,
\begin{equation}
\Qop(x)\;=\;A(x|\tau) + B(x|\tau)\;\sim\;\prod_{j=1}^{N/2} \theta_1(x-\xi_j|\tau)\;,
\end{equation}
so that the alternative Bethe variables $x_j$ and $\xi_j$ are non-trivially related.

It is interesting to note that in these exceptional cases we do not have the classification structure exposed in the previous subsection for $v<1$.

\bigskip

\noindent
\textbf{Acknowledgement.} I would like to thank Vladimir Bazhanov and Vladimir Mangazeev for valuable discussions.
Also I acknowledge the support of the Australian Research Council grant 
DP190103144.

\end{document}